\documentclass[%
twocolumn,
superscriptaddress,
nofootinbib,
showpacs,
amsmath,
amssymb,
prl,
aps
]{revtex4-1}

\usepackage{%
 microtype,
 textcomp,
 import,
 graphicx,
 dcolumn,% Align table columns on decimal point
 natbib,%
 units,
 amsmath,
 siunitx %für SI Einheiten
 }
\DeclareSIUnit\gauss{G}
\usepackage[colorlinks=false, pdfborder={0 0 0}]{hyperref}
\usepackage{color}

\definecolor{401}{RGB}{0,61,255}
\definecolor{583}{RGB}{213,143,6}
\definecolor{10.6}{RGB}{156,0,0}

\newcommand{\rem}[1]{}

\newcommand{\kB}[0]{k_{\rm{B}}}

\graphicspath{{./figures/},{.}}
\begin{document}

\title{Bose-Einstein condensation of erbium atoms in a quasielectrostatic optical dipole trap}

\author{Jens Ulitzsch}
\email{ulitzsch@iap.uni-bonn.de}
%\affiliation{Institute of Applied Physics, University of Bonn, Wegelerstr. 8, 53115 Bonn}

\author{Daniel Babik}
%\affiliation{Institute of Applied Physics, University of Bonn, Wegelerstr. 8, 53115 Bonn}

\author{Roberto Roell}
%\affiliation{Institute of Applied Physics, University of Bonn, Wegelerstr. 8, 53115 Bonn}

\author{Martin Weitz}
\affiliation{Institut f{\"{u}}r Angewandte Physik, Universit{\"{a}}t Bonn, Wegelerstrasse 8, 53115 Bonn, Germany}

\date{\today}

\begin{abstract}
Quantum gases of rare-earth elements are of interest due to the large magnetic moment of many of those elements, leading to strong dipole-dipole interactions, as well as an often nonvanishing orbital angular momentum in the electronic ground state, with prospects for long coherence time Raman manipulation, and state-dependent lattice potentials. We report on the realization of a Bose-Einstein condensate of erbium atoms in a quasielectrostatic optical dipole trap generated by a tightly focused midinfrared optical beam derived from a $\textnormal{CO}_2$ laser near 10.6\,$\mu$m  in wavelength. The quasistatic dipole trap is loaded from a magneto-optic trap operating on a narrow-line erbium laser cooling transition near \SI{583}{\nano\meter} in wavelength. Evaporative cooling within the dipole trap takes place in the presence of a magnetic field gradient to enhance the evaporative speed, and we produce spin-polarized erbium Bose-Einstein condensates with $3\times10^4$ atoms.

\end{abstract}

%\pacs{??.??.??, ??.??.??, ??.??.??}

\maketitle

The field of ultracold quantum gases was in the first years after the realization of atomic Bose-Einstein condensates limited to alkali-metal atoms almost exclusively, which have a hydrogenlike electronic structure with an $S$ ground state configuration \cite{Grimm2000}. More recently, the realization of quantum gases of atoms with more complex electronic structures, such as the rare-earth elements erbium and dysprosium, which have a nonvanishing orbital angular momentum in the electronic ground state ($L>0$), is possible \cite{Lu2011,Aikawa2012,Aikawa2014,Burdick2016a}. This opens up new opportunities for the manipulation of atoms with far-detuned laser light, as the creating of strong synthetic gauge fields \cite{Cui2013} or state-dependent optical lattices with long coherence times. Other features are a strong dipolar character. Erbium and dysprosium have magnetic moments of 7 and 10\,$\mu_\textrm{B}$, respectively, where $\mu_\textrm{B}$ denotes the Bohr magneton, with this relatively high value being due to their open $4f$ shell. The dipolar character implies novel interaction properties of such quantum gases, an issue whose consequences and possible applications have been investigated in detail in several recent works, addressing e.g. the proliferation of Feshbach resonances \cite{Frisch2014a,Baumann2014}, extensions of the Bose-Hubbard model \cite{Baier2016}, and an observed quantum ferrofluid \cite{Kadau2016}.

The dipolar character also implies that dipolar relaxation is relatively high, an issue that can be avoided by trapping the atoms in an optical dipole trap, where other than in magnetic traps atoms can be confined independently from their spin orientation. In optical dipole traps Bose-Einstein condensation of chromium, and the rare earth elements dysprosium and erbium have been achieved, and for the latter two elements with the corresponding fermionic isotopes, degenerate atomic Fermi gases also have been produced \cite{Griesmaier2005,Lu2011,Aikawa2012,Aikawa2014,Burdick2016a}. For these experiments, dipole traps realized with focused beams derived from $\textrm{Nd:YVO}_4$ or fiber lasers in the 1-1.5\,$\mu$m wavelength range were used, which is relatively far detuned from the corresponding optical transitions, though the atomic eigenfrequency is of the same order of magnitude as the optical trapping beams frequencies.

\begin{figure}[]
\includegraphics[width=0.5\textwidth]{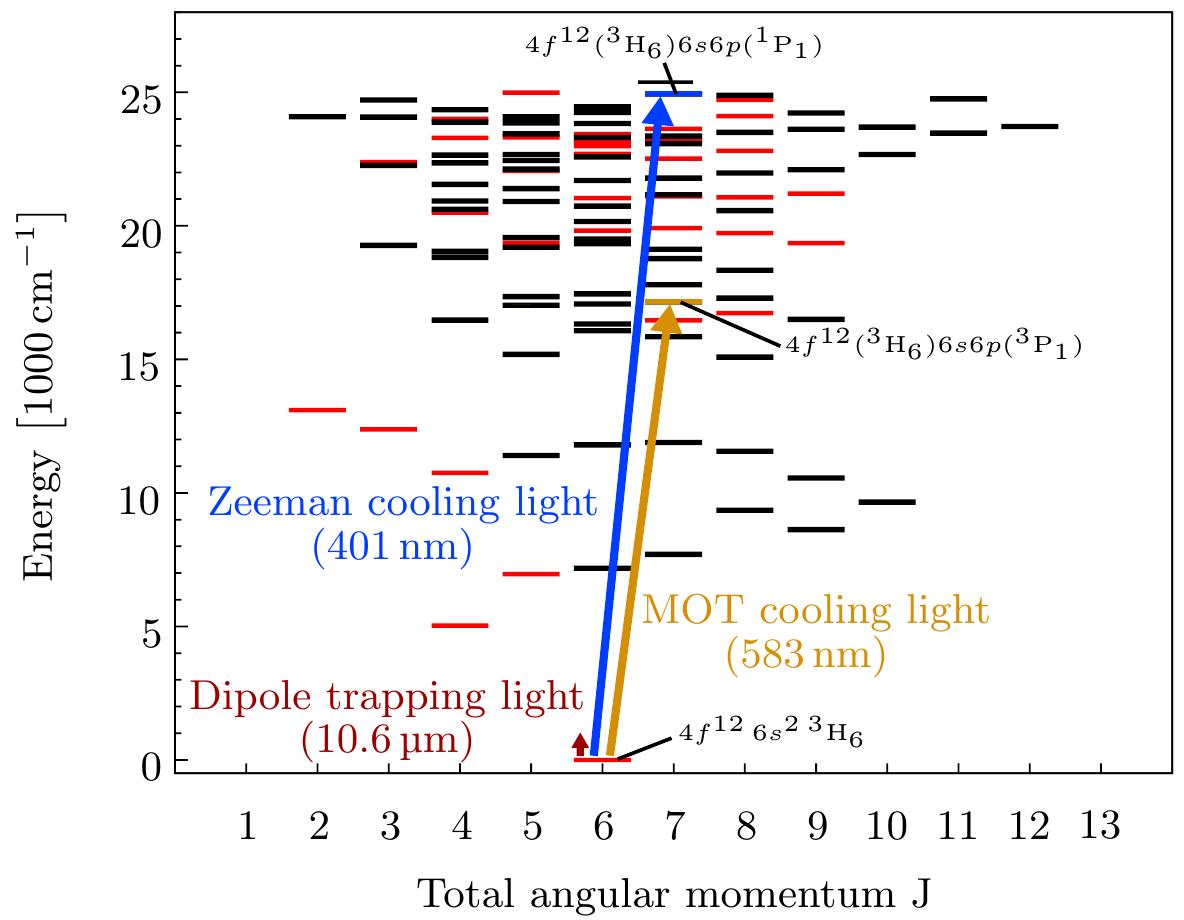}
\caption{Erbium atomic level scheme. Shown is the energetic position (in units of $hc$, the Planck constant $h$ and the speed of light $c$) of levels versus the total angular momentum $J$, where levels of negative parity are shown as red lines \cite{NISTDatabase}. For operation of the Zeeman slower, the strong cooling transition near \SI{401}{\nano\meter} is used, while the magneto-optic trap uses optical beams tuned to near the weaker electronic transition near \SI{583}{\nano\meter} in wavelength. For dipole trapping of atoms, radiation near 10.6\,$\mu$m emitted by a $\textnormal{CO}_2$ laser is used.} 
\label{fig_ErbiumEnergieschema}
\end{figure}

In contrast, the present work investigates the use of midinfrared radiation derived from a $\textnormal{CO}_2$-laser with wavelength near 10.6\,$\mu$m to induce dipole trapping of the rare earth atom erbium. $\textnormal{CO}_2$-laser dipole trapping has previously been achieved for the alkali-metal atoms rubidium, caesium, and lithium, and Bose-Einstein condensates or, for the case of the fermionic lithium isotope ${}^6\textrm{Li}$, a degenerate Fermi gas, have been successfully produced for these elements in such extremely far-off resonant traps \cite{Takekoshi,Barrett2001,Granade2002,Weber2003}. For the rare-earth element erbium, the known electronic resonances are above an energy of $hc$-5000\,$\textnormal{cm}^{-1}$ (see~fig.~\ref{fig_ErbiumEnergieschema}), corresponding to a maximum absorption wavelength of roughly 2\,$\mu$m. Since the $\textnormal{CO}_2$-laser optical emission frequency is red-detuned with respect to all known electronic resonances, we expect that atoms are pulled towards the maximum of the intensity, with the trapping potential being determined by the dynamic Stark shift:
\begin{eqnarray}
 V = - \frac{\alpha}{2} \langle |E^2| \rangle = - \frac{\alpha}{2c\epsilon_0}I,
\end{eqnarray}
where $\alpha$ denotes the quasistatic atomic polarizability, $E$ the electric field amplitude, $I$ the intensity of the trapping beam light, $c$ the speed of light, and $\epsilon_0$ the vacuum permittivity. We have used the value ${\alpha=141\times4\pi\epsilon_0\times10^{-6}\,\mathrm{C\,m}^2\,\mathrm{V}^{-1}}$ for the quasistatic polarizability of the erbium electronic ground state \cite{Lepers2014,Lepers2014a}. With an intensity ${I=2P/(\pi w_0^2) = \SI{7.4e6}{\watt\per\square\centi\meter}}$, as derived for a 63-W beam power, corresponding to the initial trapping beam power before evaporation, on a $w_0=$23\,$\mu$m Gaussian beam radius, we arrive at an expected trap depth of ${V_0\approx\SI{-2.3}{\milli\kelvin}}\times\kB$. The rate for an atom to spontaneously scatter a photon can, for a two-level system with transition frequency $\omega_0$ in the quasistatic limit $(\omega_0 \ll \omega)$ using a classical oscillator model be approximated as (from \cite{Grimm2000})
\begin{eqnarray}
	\Gamma_\textrm{sc} = \frac{2 \Gamma}{\hbar\omega_0} |V| \left( \frac{\omega}{\omega_0} \right)^3, \label{eq:Scatteringrate}
\end{eqnarray}
where $\Gamma$ denotes the upper state linewidth. 
%%%%%%%%%%%%%%%%%%%%%%%%%%%%%%
%%%%%%%%%%%%%%%%%%%%%%%%%%%%%%
The $(\omega/\omega_0)^3$ scaling of the scattering rate with the trapping beam frequency $\omega$ is a phase-space factor. The erbium atomic polarizability is in the here-relevant quasistatic limit dominated by the contribution from the strong blue laser cooling transition near \SI{401}{\nano\meter} in wavelength, so that the scattering rate can be approximated using Eq. \eqref{eq:Scatteringrate} with $\omega_0=2\pi c/\SI{401}{\nano\meter}$, $\Gamma\equiv\Gamma_\textrm{blue}\approx\SI{29.7}{\mega\hertz}$, and $V\approx-3\pi c^2 \Gamma I/\omega_0^4$. With the previously given value for the intensity at full trapping beam power, we arrive at an expected coherence time of $\tau = 1/\Gamma_\textrm{sc}$ for Rayleigh scattering of the order of \SI{1000}{\second}; see also earlier work on long trap lifetimes achieved in $\textnormal{CO}_2$-laser dipole traps with alkali-metal atoms \cite{OHara2000,Cennini2003}. In practice, in most cases the trap lifetime will be limited by the finite vacuum pressure, or in the Bose-condensed high-density case, by collisions with other trapped atoms. A technical advantage of the used midinfrared trapping wavelength is that the confocal length $b=2\pi w_0^2/\lambda$ at a given trapping beam diameter of $2w_0$ is relatively short, so that a satisfactory longitudinal atom confinement can be achieved in a single beam dipole trap, instead of the crossed beam geometry that commonly is used in all-optical Nd-laser or fiber-laser quantum gas experiments.

We here report on the trapping of the rare-earth atom erbium in a quasistatic optical dipole trap generated by a single focused $\textnormal{CO}_2$-laser beam near 10.6\,$\mu$m in wavelength. By evaporative cooling in the dipole trap in the presence of a magnetic field gradient, an atomic Bose-Einstein condensate of erbium atoms is formed. The quantum degenerate sample of rare-earth atoms in the quasielectrostatic dipole trap is an attractive sample for future studies of dipolar interaction effects and phase imprinting.
%%%%%%%%%%%%%%%%%%%%%%%%%%%%%%
\begin{figure}%[h]
\includegraphics[width=0.5\textwidth]{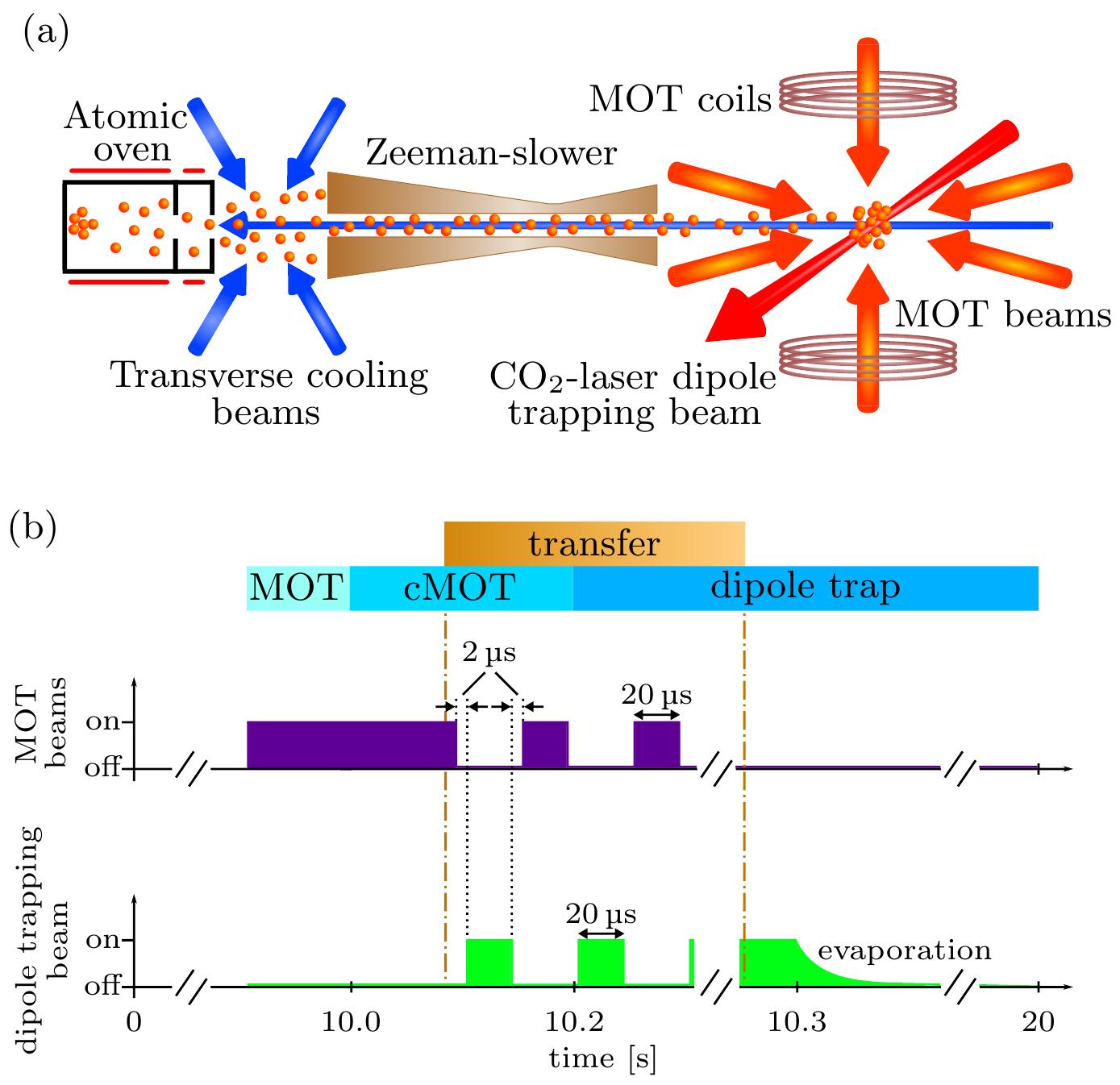}
\caption{(a) Schematic of the used experimental setup and (b) time sequence used in experimental operation. A magneto-optic trap collects cold erbium atoms from a Zeeman-slowed atomic beam. We then switch to a compressed magneto-optic trap phase, from which the quasistatic dipole trap is loaded. Atoms confined in the optical dipole trap are then evaporatively cooled to quantum degeneracy, after which the atomic cloud is analyzed with absorption imaging.} 
\label{fig_QUEST_Loading_Sequnece}
\end{figure} 
%%%%%%%%%%%%%%%%%%%%%%%%%%%%%% $4f^{12} \, 6s^2 \, {}^3\textrm{H}_6 \rightarrow 4f^{12}

A scheme of the experimental setup used to produce a Bose-Einstein condensate of erbium atoms is shown in Fig. \ref{fig_QUEST_Loading_Sequnece}(a). Like the Innsbruck erbium atom setup \cite{Aikawa2012}, the used laser cooling scheme we use is inspired from early work on Yb atoms \cite{Takasu2003,Fukuhara2007}, with deceleration of atoms in a Zeeman slower using a broad optical transition and subsequent loading of a magneto-optical trap operated using radiation resonant with a narrow-line cooling transition. An erbium atomic beam is emitted from a high-temperature effusive cell operated near \SI{1570}{\kelvin}. The beam is collimated by a two-dimensional transversal optical molasses and longitudinally decelerated by a spin-flip Zeeman slower. These stages use light tuned to near the strong $4f^{12}6s^2({}^3\textrm{H}_6) \rightarrow 4f^{12}({}^3\textrm{H}_6)6s6p({}^1\textrm{P}_1)$ blue cooling transition near 401-nm wavelength of $\Gamma_\textrm{blue} \approx \SI{29.7}{\mega\hertz}$ natural linewidth. The cooling light is derived from the emission of a frequency-doubled Ti:sapphire laser with \SI{1.2}{\watt} of power. The laser system frequency is locked to the blue erbium cooling transition by means of Doppler-free modulation transfer spectroscopy in a hollow cathode discharge cell \cite{McCarron2008,Brammer2012}. The emission of the frequency-doubled source is split up to provide the required optical beams for transverse cooling and Zeeman slowing of the atomic beam, respectively, and also for absorption imaging of the trapped atomic cloud, while frequency shifting is in all cases accomplished with acousto-optic modulators. To collect a laser-cooled atomic cloud of ${}^{168}\textrm{Er}$ atoms, we use a narrow-line magneto-optic trap, operated with laser beams tuned to the red of the $4f^{12}6s^2({}^3\textrm{H}_6) \rightarrow 4f^{12}({}^3\textrm{H}_6)6s6p({}^3\textrm{P}_1)$ transition of the erbium atom near \SI{583}{\nano\meter} (natural linewidth $\Gamma_\textrm{yellow}\approx\SI{186}{\kilo\hertz}$). As a light source for the magneto-optic trap, we use a dye-laser system, which is frequency stabilized by locking to an external ultralow expansion cavity via the Pound-Drever-Hall technique \cite{Drever1983}, providing a linewidth on the order of \SI{25}{\kilo\hertz} relative to the reference cavity. The typical frequency drift of the reference cavity is \SI{10}{\kilo\hertz\per\day}, which is low enough to allow stable operation of the experiment without active frequency stabilization to an atomic standard. The magneto-optical trap is operated with three orthogonal, retroreflected optical cooling beams. The beams spatial profiles are relatively flat, as an aperture is used to clip the outer parts of the Gaussian wings. The diameter of the formed cooling beams is \SI{35}{\milli\meter}.

During the 10-s loading phase of the magneto-optic trap, we operate with a relatively large detuning of $-34\,\Gamma_\textrm{yellow}$. At this time, the competition of gravity and the weak optical cooling force leads to a $\sim\SI{1}{\centi\meter}$ spatial shift down relative to the center of the magnetic quadrupole field \cite{Takasu2003,Loftus2004,Fukuhara2007}. At the spatially shifted position of the atomic cloud at this loading time the intensity of the blue Zeeman cooling beam is low, which reduces the loss rate from the finite leakage of the 401nm erbium transition. To prepare loading into the optical dipole trap, we compress the magneto-optic trap by reducing the detuning of the cooling beams to $-2.7\,\Gamma_\textrm{yellow}$. Simultaneously, the intensity of the MOT cooling beams is reduced to $2\,I_\textnormal{S}$, and the Zeeman cooling light is extinguished. The reduction of the MOT beams detuning besides causing a compression also spatially moves the atomic cloud towards the center of the magnetic quadrupole field. The typical number of atoms in the compressed MOT is $2.5 \times 10^7$ at a temperature of 20\,$\mu$K. This temperature value is clearly below the expected depth of the trapping potential in the quasi-static optical dipole trap.

As a light source for dipole trapping of erbium atoms, we use a single frequency $\textnormal{CO}_2$ laser with 120-W nominal output power, and the emitted beam intensity is controlled via an acousto-optic modulator. The beam enters the vacuum chamber through a ZnSe viewport, and is then focused with an adjustable ZnSe lens to a Gaussian beam radius of 23\,$\mu$m, spatially overlapping with the position of the atomic ensemble trapped in the compressed magneto-optical trap. For an analysis of the trapped atomic ensemble confined in the optical dipole potential after the dipole trapping phase, the midinfrared beam is extinguished and a standard absorption imaging technique with a weak beam resonant with the blue 401-nm erbium cooling transition is applied. We experimentally do not observe a transfer of atoms into the dipole trap when activating the $\textnormal{CO}_2$-laser dipole trapping beam already during the MOT phase. This is attributed to the ac Stark shifts in the upper and lower states of the cooling transition differing from each other, which presumably leads to a blue detuning within the dipole beam focus and correspondingly to a heating of atoms in this region. We note that this is in contrast to observations reported from closer resonant erbium dipole trapping experiments \cite{Aikawa2012}.
%%%%%%%%%%%%%%%%%%%%%%%%%%%%%%
\begin{figure}[t]
\includegraphics[width=0.5\textwidth]{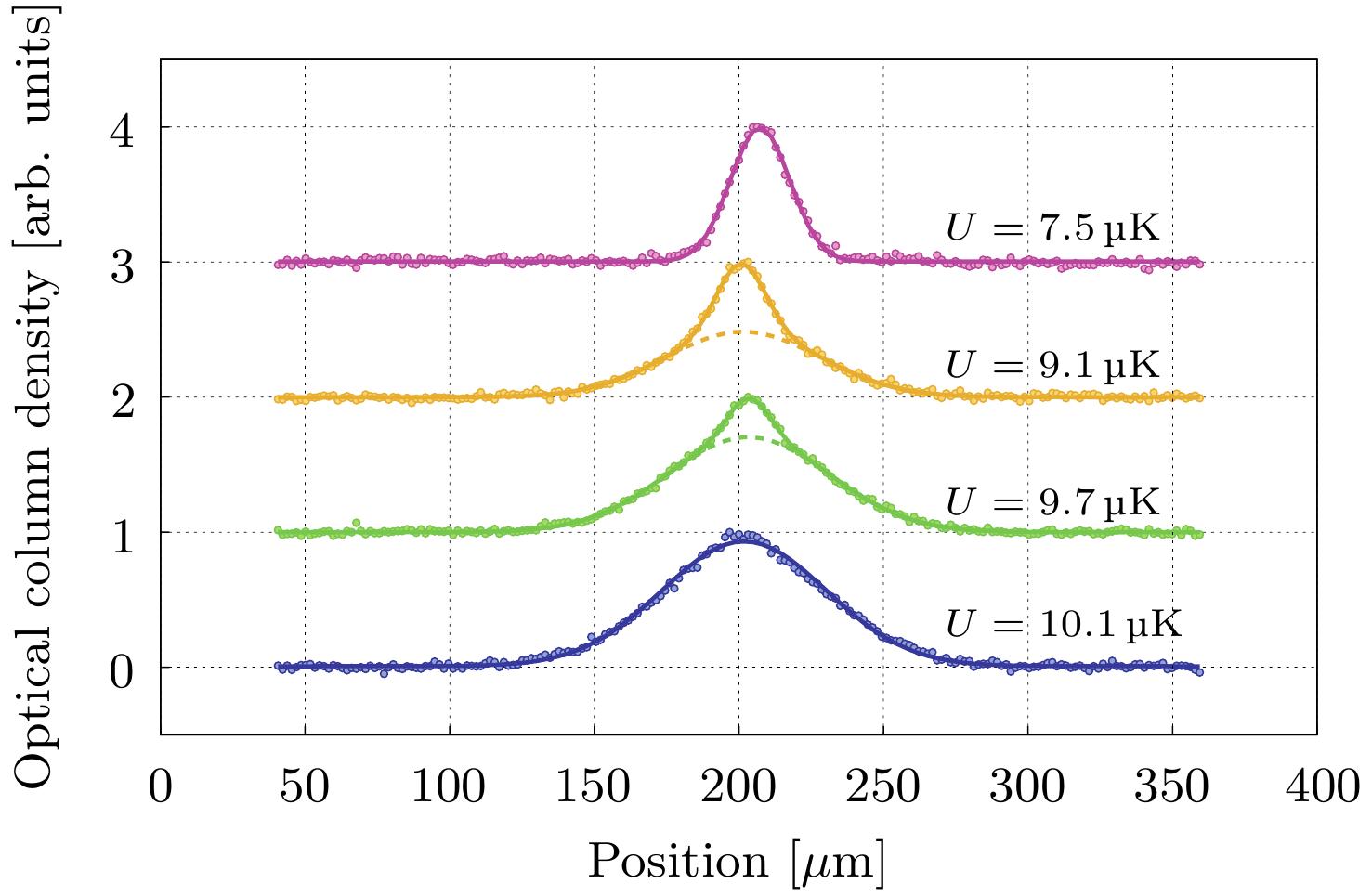}
\caption{Optical column density (vertically shifted) of cold erbium atomic clouds versus transverse position, as measured with a weak optical probe pulse tuned to the 401-nm erbium transition, for several values of the final trapping beam power. As the depth of the dipole trap decreases (values for trap depth refer to calculated values in units of the Boltzmann constant, $k_B$), the distribution becomes bimodal, corresponding to a Bose-Einstein-condensed peak on top of a broad thermal cloud. The topmost plot gives data for an almost pure Bose-Einstein condensate.} 
\label{fig_Bimodal}
\end{figure} 
%%%%%%%%%%%%%%%%%%%%%%%%%%%%%%
In our experiment, successful loading of atoms into the dipole trap was observed when activating the dipole trapping beam only after extinguishing of the MOT cooling beams' radiation, and the observed transfer efficiency could be enhanced by repeatedly alternating between activating the cooling beams and the dipole trapping beam, respectively, during loading of the quasistatic dipole trap \cite{Dalibard1983,Chu1986}. With the used sequence of alternating between 20-$\mu$s-long cooling and dipole trapping potentials, respectively [see Fig. \ref{fig_QUEST_Loading_Sequnece}(b)], resulting in a time-averaged dipole potential during which cooling is possible, we in a 11-ms-long loading phase observe a factor 2 increase of the transfer efficiency.

We typically transfer $1 \times 10^6$ erbium atoms into the quasistatic dipole trap. The measured radial and longitudinal trap frequencies at this point are \SI{2.6}{\kilo\hertz} and \SI{260}{\hertz}, respectively. The atomic temperature, as determined with a time-of-flight measurement, is 177\,$\mu$K, and the atomic phase-space density can be calculated to $\textrm{PSD}_\textrm{QUEST}\approx0.37 \times 10^{-4}$.
%%%%%%%%%%%%%%%%%%%%%%%%%%%%%%
\begin{figure}[b]
\includegraphics[width=0.5\textwidth]{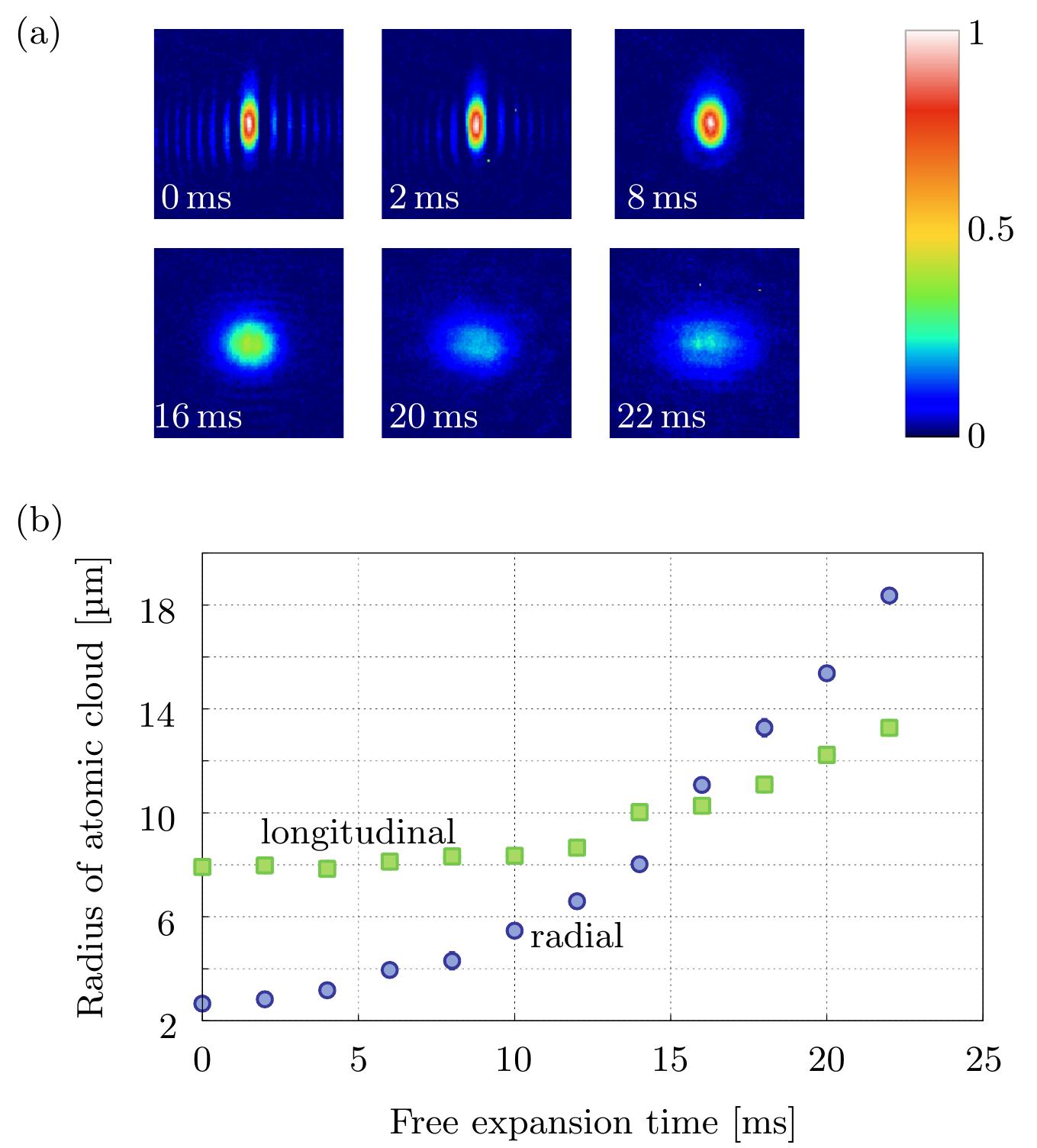}
\caption{(a) Shadow images of the atomic erbium clouds for different expansion times after release from the dipole trap. The color bar shows the optical column density in arbitrary units. (b) Variation of the radial (blue dots) and longitudinal (green squares) 1/$e{}^2$ width versus the time of free flight, showing the expansion of the atomic cloud upon release from the trap. At approximately 16-ms free expansion time, the aspect ratio inverts.} 
\label{fig_Rinversion}
\end{figure}
%%%%%%%%%%%%%%%%%%%%%%%%%%%%%%

To cool atoms towards quantum degeneracy, forced evaporative cooling is applied. Within a 10-s-long phase, the dipole trapping beam power is ramped down to approximately \SI{0.8}{\percent} of its initial power, allowing for higher energetic atoms to escape, with the remaining atoms by elastic collisions rethermalizing to lower temperature distributions. The evaporation occurs in the presence of a magnetic field gradient, which is increased from \SI{1.7}{\gauss\per\centi\meter} to \SI{3.8}{\gauss\per\centi\meter} during the ramp of the dipole trapping beam intensity \cite{Hung2008}. We found that the application of such a magnetic field gradient, a technique expected to be particular useful for atoms with large magnetic moment, reduces the required time for the evaporative cooling phase, as attributed to the corresponding Stern-Gerlach force (which is codirected to the direction of gravity) aiding high energetic atoms to leave the trap. We typically produce spin-polarized Bose-Einstein condensates with $3\times10^4$ erbium atoms of the isotope ${}^{168}\textrm{Er}$.

To validate the cross-over to quantum degeneracy, the spatial distribution of the atomic ensemble is investigated by absorption imaging. Figure \ref{fig_Bimodal} shows experimental data for the measurement optical density of the atomic cloud as a function of the position in a line cut of the images for different effective trap depths after evaporation. The lowest profile shown was recorded at a trap depth where the distribution is still thermal, while for the next two profiles the trapping beam power was ramped down to values where the distribution is bimodal, with a condensate peak on top of the thermal cloud. The top profile shown, recorded at the lowest final trap depth, resulted in an almost pure Bose-Einstein condensate. The critical temperature below which we achieve condensation is approximately \SI{100}{\nano\kelvin}.

%%%%%%%%%%%%%%%%%%%%%%%%%%%%%%
\begin{figure}[t]
\includegraphics[width=0.5\textwidth]{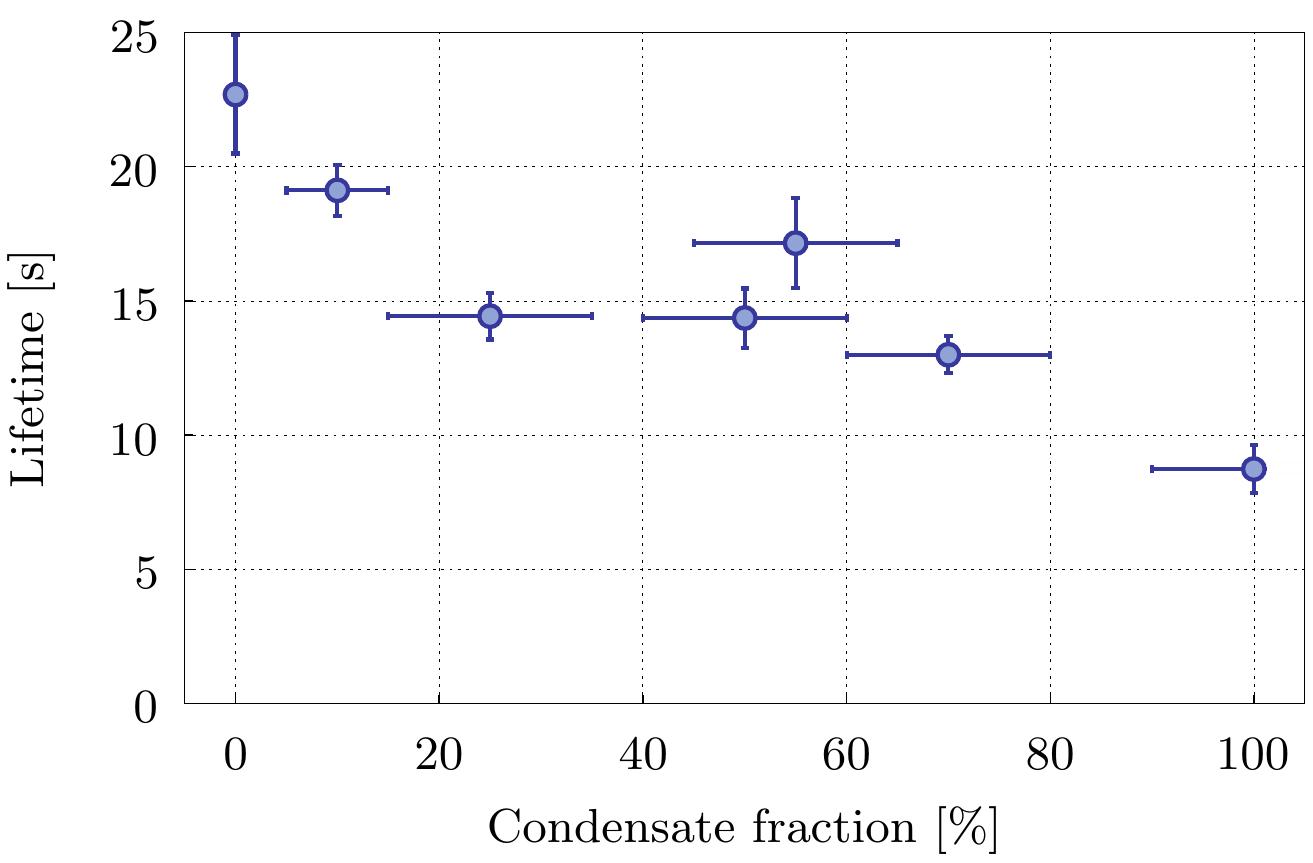}
\caption{Measured lifetime of atomic clouds confined in the quasistatic optical dipole trap versus the condensate fraction. The leftmost data point was recorded for a thermal cloud with temperature slightly above the BEC transition temperature. As the condensate fraction increases, the observed lifetime reduces.} 
\label{fig_KLebensdauer}
\end{figure}
%%%%%%%%%%%%%%%%%%%%%%%%%%%%%%
Figure \ref{fig_Rinversion}(a) gives absorption images recorded at different times after releasing the atoms from the trap, and figure \ref{fig_Rinversion}(b) the 1/$e{}^2$ radius both in the radial and in the longitudinal directions as a function of the free expansion time. One observes here the expected inversion of the aspect ratio, as the cloud during the expansion evolves from a prolate to an axial geometry. Finally, Fig. \ref{fig_KLebensdauer} gives the measured 1/$e$ lifetime of the atomic cloud confined in the dipole trap versus the condensate fraction. As the condensate fraction increases, the trap lifetime decreases due to interaction effects, and reaches \SI{8.7}{\second} for an almost pure condensate. We attribute the observed trap lifetime at small condensate fractions to be mainly vacuum limited (in addition to residual free evaporation of atoms occurring).

To conclude, we have demonstrated the quasistatic optical dipole trapping of an ensemble of rare-earth atoms. An atomic erbium Bose-Einstein condensate was produced within the dipole trapping potential induced by a single, tightly focused optical beam emitted by a $\textnormal{CO}_2$ laser operating near 10.6\,$\mu$m wavelength. The application of a moderate magnetic field gradient during the evaporation phase for the large magnetic moment erbium atom was found to enhance the evaporation speed.

To further enhance the number of produced condensate atoms, a broadening of the MOT beams' frequency \cite{Maier2014,Dreon2017}, ``dark-state'' cooling techniques within the dipole trapping potential \cite{Boiron1998}, and, spatial modulation techniques of the dipole trapping beam to increase the overlap of the MOT with the dipole trap volume \cite{Han2001,Kinoshita2005} could be investigated. For the future, ultracold atomic erbium samples are promising systems for both the exploration of novel states of matter, taking advantage of the strong dipole-dipole interaction \cite{Frisch2014a,Baumann2014,Baier2016,Kadau2016} and long coherence time Raman manipulation due to the nonvanishing orbital angular momentum of this atom in the electronic ground state \cite{Cui2013}. 
\\
%%%%%%%%%%%%%%%%%%%%%%%%%%%%%%%%
%    Acknowledgments
%%%%%%%%%%%%%%%%%%%%%%%%%%%%%%%%
\begin{acknowledgments}
We thank H. Brammer for his contributions during the early phase of the experiment. Financial support of the DFG within CRC-TR 185 is acknowledged.
\end{acknowledgments}

%\bibliography{library_2017_02_22}

%%%%%%%%%%%%%%%%%%%%%%%%%%%%%%%%%%%%%%%%%%%%%

\end{document}